\documentclass[12pt]{article}

\usepackage{graphicx}
\usepackage{times}
\usepackage{xcolor}

\newenvironment{myabstract}{%
\begin{quote} \bf}
{\end{quote}}

\title{ On Behind the Physics of the Thermoelectricity of Topological Insulators }

\author
{Daniel Baldomir$^{1\ast}$ and Daniel Fa\'ilde$^{1}$\\
\\
\normalsize{$^{1}$Departamento de F\'isica Aplicada, Instituto de Investigaci\'ons Tecnol\'oxicas. Universidade de} \\
\normalsize{Santiago de Compostela E-15782 Campus Vida s/n, Santiago de Compostela, Spain}\\
\\
\normalsize{$^\ast$daniel.baldomir@usc.es}
}

\date{}

\begin{document}

\baselineskip24pt

\maketitle

\begin{myabstract}
  Topological Insulators are the best thermoelectric materials involving a sophisticated physics beyond their solid state and electronic structure. We show that exists a topological contribution to the thermoelectric effect that arise between topological and thermal quantum field theories applied at very low energies. This formalism provides us with a quantized topological mass proportional to the temperature T, being both quantities directly related with an electric potential V and getting a Seebeck coefficient where we identify an anomalous contribution that we associate to the creation of real electron-hole Schwinger's pairs close to the topological bands. Finally, we find a general expression, considering the electronic contribution, for the dimensionless figure of merit of these topological materials, getting a value of 2.73 that is applicable to the Bi$_2$Te$_3$, for which it was reported a value of 2.4, using only the most basic topological numbers (0 or 1).
\end{myabstract}

\section*{Introduction}

Nowadays topological insulators (TI) are the best thermoelectrics (TE) at room temperature,\cite{Hasan,Chang,Ryuji,Xu,Reza} specially if they are combined with nanotechnological structures which are able to reduce the phononic thermal conductivity. A good example is\cite{Wright} the bismuth telluride, Bi$_2$Te$_3$, that has a small band gap giving a good number of carriers at room temperature (300K) and reaching 2.4 for its dimensionless figure of merit ZT for p-type using alternating layers in a superlattice with $Sb_2Te_3$. This is the highest value of thermoelectricity ever observed \cite{ven} so far at room temperature. Despite the fact that the electronic structure of these materials was exhaustively studied  \cite{esch}  in relation with their thermoelectricity\cite{Müchler,Lado}, there is still lacking in the literature a physical model\cite{Aydemir} which could take explicitly into account their common topological and physical features. Its importance might appears obvious with a counterexample, the Pb$_{1-x}$Sn$_x$Te has a good electronic structure to be a topological insulator with thermoelectricity \cite{PbSnTe}, but due to have an even number of band inversions, this prevents it to have time reversal symmetry $\hat{T}$ and thus to be a TI, although it is a good thermoelectric at higher temperatures. Hence, the whole topology of a TI is not fully necessary for having good thermoelectricity as we are going to see. Understanding the physics behind these phenomena is not an easy task because links scientific branches which were developed independently: particle physics, statistical mechanics, condensed matter and algebraic topology. This is a characteristic of materials which exhibit linear dispersion laws instead of quadratic ones, allowing a quantum field interpretation where the spinors play a fundamental role substituting the usual non relativistic wave function. 

We organize the paper as follows. We examine the basic concepts of topology and physics for topological insulators trying to show how they are related in a same structure. After that we show that the Riemann-Hurwitz formula plays an important role which was not considered in the literature so far. This allows us to find five topological regions which are connected by four bands  defined using the periodicity of the instanton solutions associated to the non-Abelian Berry fields introduced within the bulk of the TI. In the case that there is not time reversal symmetry $\hat{T}$ we have only three topological regions, connected by only two bands. Finally we present a straightforward relationship between the temperature T and the topological index $\mu$ with the scalar electric potential V. This leads to a Seebeck coefficient for which we identify two terms, one related to the topological electron pump, and another associated to a change in the topological index that might be associated to the creation of real electron-hole pairs as we will analyze. We end with the calculation of a general expression for the edge states topological figure of merit in TI, taking into account the electronic contribution and neglecting the phononic part \cite{ven,holey}.

Solid state physics allow us to tackle the problem of a crystal, with translation symmetry, reducing the analysis of the different physical properties to the first Brillouin zone. This is permitted thanks to Bloch theorem, i.e. for a periodic potential $V(x+a)=V(x)$, where $a$ is the spatial period or the lattice constant, the wave function associated to the electrons have also a periodicity $\psi_k(x+a)= e^{ika}\psi_k(x)$. Being the eigenvalues $\xi_k=\xi_{k+T}$,$\xi_k=\xi_{-k}$ also periodic in time. These are the necessary conditions for calculating the energy bands in a solid. It is easy to see that parity $\hat{P}$, or space inversion, is straightforwardly followed, while time reversal symmetry $\hat{T}$ is not so obviously fulfilled. For example, the Schr\"odinger equation $i\hbar \frac{\partial}{\partial t} \psi(t) =H \psi(t)$ under time reversal gives $i\hbar \frac{\partial}{\partial (-t)} \psi(-t) =H \psi(-t)$, where $\psi(-t)$ is not a solution due to the first order time derivative. This can be solved if the $\hat{T}$ operator has also associated a complex conjugation $\hat{K}$ operator. In fact, we must define $\hat{T}=U\hat{K}$ for spinless states where $U$ is a unitary operator. In the case of having half integer spin particles, the unitary operator can be written in function of the $\sigma_y$ Pauli matrix as $U=exp({-i\frac{\pi}{2}\sigma_y})$, which for spin $1/2$, given that $\sigma_y^2=1$, allow us to write the time reversal operator as $\hat{T}=-i\sigma_y \hat{K}$, which acting on the multiparticle state gives $T^2=1$ for an even number of fermions or $T^2=-1$ when the number is odd. More generally written, $T^2$ has eigenvalue $(-1)^{2s}$ for a particle of spin $s$, and if $|n>$ is an energy eigenstate, then $T|n>$ too, sharing the same energy and being orthogonal to each other. Thus, if there is an odd number of electrons there must be (at least) a twofold degeneracy; this is known as Kramers degeneracy. This is Kramers theorem, which when it is completed with the previous Bloch theorem for the bands, should provide us the first tools\cite{Fu} to examine topologically the first Brillouin Zone (BZ) of a TI. 
  
In Fig.1 we represent schematically how the 2D BZ evolves under different translation operations transforming the square BZ into two equivalent cylinders $S_1$. Consequently, is also easy to see that combining the two lattice periodicities available in two dimensions, a 2D torus $S^1 \times S^1$ can arises. That is, the existence of translation symmetry allow us to transform a topologically trivial square into a non-trivial torus with genus $g\neq0$. Given that all crystals are able to develop these fundamental properties, non-trivial topology is not enough when we study TI and the presence of singularities which are usually counted on the band structure are needed. Obviously, without extra-information, these singularities break the translation symmetry or the periodicity of the Bloch states \cite{Haldane,Kane}. So, let's analyze a little bit the possibility of these translations symmetry breaking. Physically the infinitesimal translations are generated by the linear momentum operator $ \hat L(dx)=\hat1-\frac{i}{\hbar} \hat p(dx)$, where $[\hat x,\hat p]=i \hbar \hat 1$  follows the Heisenberg's principle of uncertainty. Extending this operation over the whole crystal one gets the finite translation operator $ \hat L= \lim_{N \rightarrow \infty} [\hat1-\frac{i}{\hbar} \hat p (\frac{a}{N})]^N =\exp (-\frac{i}{\hbar} \hat p a)$. Taking into account the periodicity of the crystal potential $V(x)=V(x+a)$, which involves the periodicity of the states, we can write $L_a|\psi(x)>=|\psi(x-a)>=\exp(ika) |\psi(x)>$   where $k$ is the wavelength number associated the momentum $p=\hbar k$. On the contrary, the presence of singularities make that $|\psi(x-a)>=\exp(ika) \exp(i \gamma) |\psi(x)>$, being $\gamma$ the Berry phase, giving to electrons an anomalous phase factor during a complete cycle in the order parameter. Thus, in order to maintain the periodicity of the Bloch states, preserving the non-trivial topology that results for having a torus, singularities must take place on the edges of the crystal, where translation symmetry is no longer satisfied. Since singularities manifest in energy bands by means of Fermi points, bands are good hosts for feeding the non trivial topology of a material whose electronic structure is appropriate and which its surface offers the breaking of translation symmetry, besides other well known fields features for lower dimensions. In this way, the Berry phase joints the non-trivial topology of the crystal employing its curvature and connection on the bands\cite{Sim} as we are going to see.

\begin{figure}[h]
\begin{center}
\includegraphics[scale=0.4]{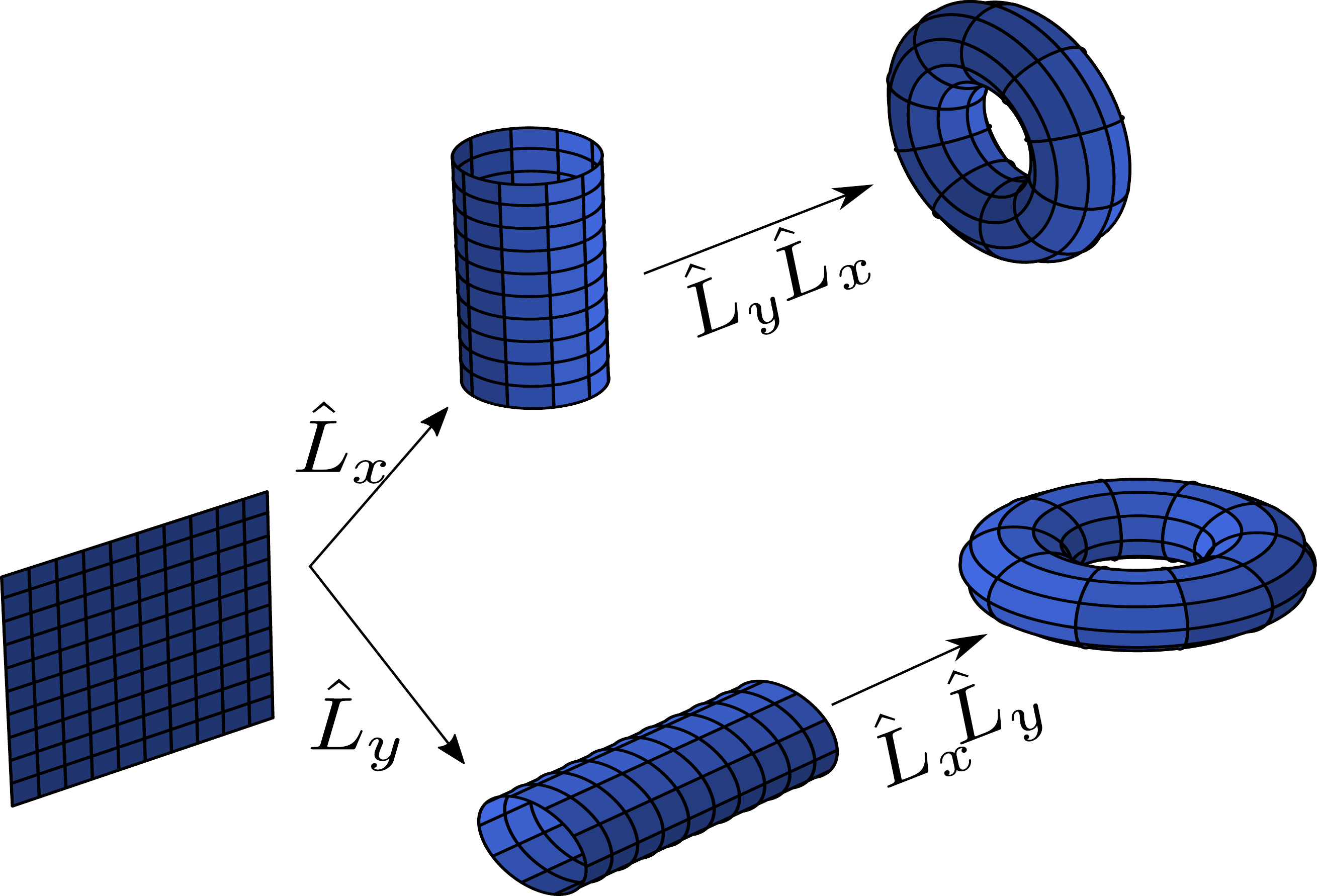}
\caption{\textbf{Evolution of the $2-D$ Brillouin Zone with trivial topology under different translation symmetries}. Applying individual translation symmetries, two cylinders equivalent to two non trivial topological $S^1$ circles arise. Combining the two traslations symmetries we obtain two equivalent torus $T^2=S^1\times S^1$.}
\end{center}
\end{figure}

The topological elements of the TI have been found, but now it remains to see how they work together employing their associated invariants. The Riemann-Hurwitz formula \cite{Hartshorne}, which generalizes the Euler topological invariant, enable us to to construct an equation relating the genus g and g' of two compact surfaces, i.e., whose boundary is zero. Actually this formula establishes the conditions for a map $ f: T {\rightarrow} S $ being subjective and holomorphic, reducing the several topological invariants introducing the genus of $T$, the genus of $S$, $N$ the degree of the map $f$ and the amount of ramifications $e_{f}(k)$. Riemann used the mentioned formula in the case that $S$  were zero, i.e., spheres. Much later the proofs were given by Zeuthen and Hurwitz. The formula is 

\begin{equation}
       2(g_{T} -1)=2N(g_{S} -1)+\sum_{k\in T} (e_{f}(k)-1)
\end{equation}

In our case we have a 2D torus $T^2$ with genus $g_T=1$ mapped in a $S^3$ sphere with genus $g_S=0$ and the degree of the map $N=2$ due to the Kramers double degeneracy, see Fig. 2. Hence we obtain five ramifications branches, i.e., $e_f(k)=5$. This can be directly related with the second Chern number as we shall see later, playing a fundamental role in the transformation of heat in electricity. The number of ramification branches diminish to three when there are not Kramer pairs, that is, $N=1$ with $e_f(k)=3$. But it is very remarkable to observe that this formula doesn't depend of the dimensions of the involved sphere or torus, which justifies us to work with a $T^2$ torus instead of a $T^3$ or $T^4$ without being worry about new results.

\begin{figure}[h]
\begin{center}
\includegraphics[scale=0.5]{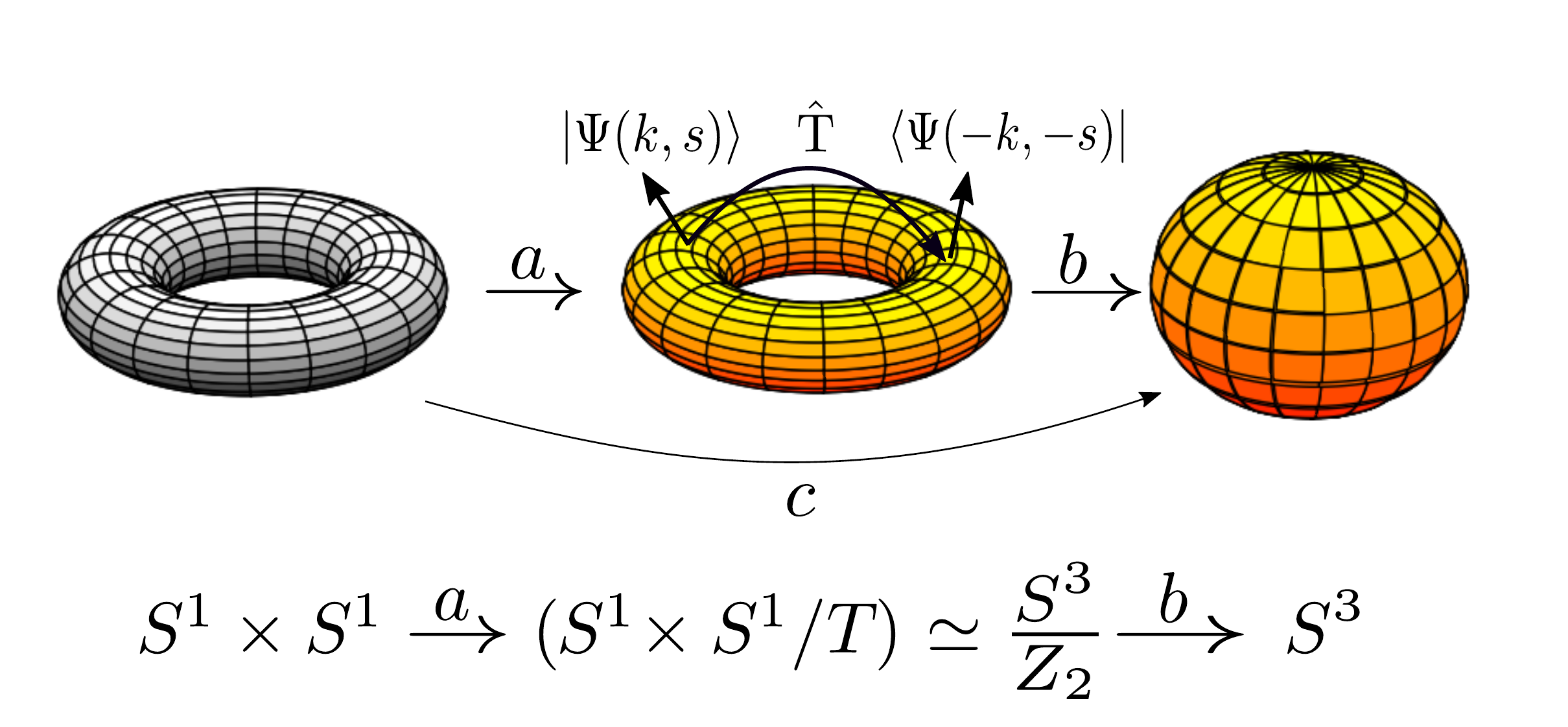}
\caption{\textbf{Schematic illustration  of the projection of a torus into a sphere}. The first torus is mapped by a on the second torus, which is equivalent to one sphere since it includes Kramers pairs with the same energy and orthonormal. The map b carries points of the previous torus on a real sphere. It is inmediate to see that the map $c=b \cdot a $ is conmutative.}
\end{center}
\end{figure}
 
In the presence of singularities in the band structure the map c of Fig. 2 can be interpreted as the $d_a$ map within the Hamiltonian $H$ introduced\cite{Xiao} to study TI in (4+1)D,
\begin{equation}
H=\sum_{k} \psi_k ^+ d_a (k) \Gamma^a \psi_k
\end{equation}
being $d_a (k)=\left(m+p\sum_i \cos k_i,\sin k_x, \sin k_y, \sin k_z, \sin k_w\right)$ and $\Gamma's$ the Clifford matrices $\{\Gamma^\mu , \Gamma^\nu \}= 2 g^{\mu \nu} 1_{5 \times 5}$. We can calculate then, the second Chern number $C_2$ associated to this Hamiltonian
\begin{equation}
C_2 = \frac{3}{8 \pi^2} \epsilon^{abcde} \int{dk^4 } \hat{d}_a (k) \frac{\partial \hat{d}_b (k) }{\partial k_x} \frac{\partial \hat{d}_c (k) }{\partial k_y} \frac{\partial \hat{d}_d (k) }{\partial k_z} \frac{\partial \hat{d}_e (k) }{\partial k_w}
\end{equation}
which is no more than the winding number resulting from the map $\hat{d_a(k)}\equiv\frac{d_a(k)}{|d(k)|}$  of a four dimensional torus $T^4$ to a sphere $S^4$, having the mass $m$ associated to the spinor's $\Gamma's$ five critical values  given by the condition $\sum_{k,\mu} d_a^2=0$ allowing us to identify, in the same way as the Riemann-Hurwitz theorem, five different branches for the second Chern number
\begin{equation}
C_2 \equiv \mu=\left\{ \begin{array}{ll}  0  ,  \textrm{$ m \notin (-4p,4p)$}\\
  +1 ,  \textrm{$ m \in (-4p,-2p)$}\\
	-3 ,  \textrm{$ m \in (-2p,0)$}\\
	+3 ,  \textrm{$m \in (0,2p)$}\\
	-1 ,  \textrm{$ m \in (2p,4p)$}\\
	\end{array} \right.
\label{5ram}
\end{equation} 
where the parameter p is taken as $p=\frac{2k_BT}{v^2_F}$  and must be equal to the background kinetic energy of the particles for keeping its physical meaning. This parameter $p=\frac{2k_BT}{v_F^2}$ is chosen as a mass parameter that is mainly associated to the quantization of the temperature\cite{Deser}, as we are going to see very soon. It is easy to see that for non-$\hat{T}$ symmetric Hamiltonians there are only three ramifications available for the Second Chern number, in the same way as it happens when we break time reversal symmetry in a Quantum Spin Hall system. As we saw, this can be interpreted, in one straighforward form, into the Riemann-Hurwitz formalism as two different maps, where we change the degree of the map N from 2 to 1. 

Physically the non-Abelian Berry phase takes into account the allowed bulk degenerate states which are directly related with the change of temperature, as we shall see soon. First, we take the definition of the non-Abelian Berry connection $a^{\alpha \beta}_j(k)=i<\alpha,\textbf{k}|\frac{\partial}{\partial_{k_j}}|\beta,\textbf{k}>$ and the associated field (or curvature) $f^{\alpha \beta}_{ij}=\partial_i a^{\alpha \beta}_j - \partial_j a^{\alpha \beta}_i + i[a_i,a_j]^{\alpha \beta}$ and second Chern number $C_2=\frac{1}{32\pi^2} \int d^4k \epsilon^{ijmn} tr(f_{ij} f_{mn})$, where the indixes of the Levi-Civita tensor stand by $i,j,m,n=1,2,3,4$ and $\alpha$ refers to the occupied bands. This can be written within a pure gauge Yang-Mills formalism, where its solutions transform in general by $a_\mu=T^i a^i_{\mu}$ and $f_{\alpha \beta}=T^i f_{\alpha \beta}^i$ being the $T^i$'$s$ the generators of the inner symmetry group, which in our case will be $SU(2)$ \cite{nakahara2003geometry}. For a $U \in SU(2)$ being position dependent, we have the gauge transformations for the potentials $a_\alpha {\rightarrow} a_\alpha^U=U^{-1}a_\alpha U + U^{-1} \partial_\alpha U$ and the fields $f_{\alpha \beta} {\rightarrow} f_{\alpha \beta}^U=U^{-1} f_{\alpha \beta} U$. It is immediate to observe that the non-Abelian fields are not gauge invariant, at difference of what happens with the Abelian ones, which tell us that the curvature depends of the group representation but not the topological numbers. On the other hand the Lorentz invariant associated to them, $\epsilon^{ijmn} f_{ij}f_{mn}$, classically doesn't appear in the action because it is a surface term negligible under variations, but quantum mechanically it must be introduced due to have the action on an exponent and being this term which accounts basically for the topology of the motion equations where they are defined. Thus, given that $tr(\epsilon^{ijmn} f_{ij}f_{mn})=4\epsilon^{ijmn} \partial_i[tr(a_j\partial_m a_n-\frac{2}{3}a_j a_m a_n)]$ this allows to define the Chern-Simons term $W(a)=-\frac{\mu}{8\pi^2}\int d^3x \epsilon^{ijm} tr(a_i\partial_j a_m-\frac{2}{3}a_i a_j a_m)$ where $\mu$ is a topological mass, which is not gauge invariant under gauge transformations 

\begin{equation}
W(a^U) {\rightarrow} W(a)-\frac{\mu}{8\pi^2} \int d^3x[ \epsilon^{ijm}
\partial_i tr(\partial_j U U^{-1} a_m)+\frac{\mu}{24 \pi^2} \int d^3x \epsilon^{ijm}tr(U^{-1}\partial_i U U^{-1}\partial_j U U^{-1} \partial_m U)]
\end{equation}

The first term of the integral is a total derivative which can be made zero in a manifold without boundary, while the second, is an integral written in short as $w(U)$, which provide us topological information of the manifold as a winding number. This integral is actually an integer number coming from the homotopy group $\pi_3(S^3)$ for the $SU(2)$ group and we can take the  Abelian Chern-Simons gauge symmetry associate the background electrodynamics, i.e., the U(1) as a subgroup of transformations close to the surface. In this case, we can rewrite the Chern-Simons transformation for both fields. Hence the transformation of the Chern-Simons can be rewritten as 
\begin{equation}
W(a) {\rightarrow} W(a)+ \frac{\mu}{12\pi} 24\pi^2 w(U)=W(U)+2\pi\mu w(U)
\end{equation}

But as the path integral $e^{i\frac{W}{\hbar}}$ must be invariant, hence it means that $\mu$ is an integer. These are going to be the dynamic fields meanwhile the electromagnetic gauge potential $A_\alpha$ will appear as a background gauge field. In fact the electric field is due to the electric polarization $P$ associated to the cooperative Berry phase within the energy bands of the material, whose topological origin is fundamental.

Let us go now to the thermodynamic part of the topological thermoelectricity, applying the instanton solution associated to the second Chern number developed above. Within this mathematical contest it is possible to make a direct relationship between thermodynamics and quantum formalism. Feynman path integral give us an expression for the amplitude of probability to evolve a particle from $x_i(0)$ to $x_f(t)$ a time later by $<x_f|e^{-iHt}|x_i>=\int^{x_f}_{x_i}{Dx  e^{-iS}}$, where $S$ is the classical action, which could be obtained by $S=\int^t_0 dt'{[\frac{p^2}{2m}-V(x)]}$. On the other hand, in Statistical Mechanics the partition function is defined in quite a similar form by $Z(\beta)=tr e^{-\beta H}$ where $\beta \equiv \frac{1}{k_BT}$. There is a way to change from one to the other using the Wick trick, i.e., doing the time a pure complex variable and transforming the Minkowskian space-time in a Euclidean one. Thus, we have $-iS=\int^{-i\tau}_{0}{d\tau'[\frac{p^2}{2m}-V(x)]}=S_E$ where $S_E$ is the new Euclidean action. This fixes the concept of temperature relating it directly with the time by $\tau=\hbar \beta$ making $tr e^{-\beta H}=\int{dx_i <x_i|e^{-\beta H}|x_i>}=\int{dx_i}\int^{x(\beta)=x_i}_{x(0)=x_i}{Dx e^{-S_E}}$, where we have assumed a cyclic motion $x(0)=x(\beta)$ i.e. the particle must come back where it started after Euclidean time $\tau=\hbar \beta$,\cite{M,D}. This is exactly how the Berry's phase works and justify partially its introduction as a gauge field.

On the other hand, Berry's phase is directly related with the electric polarization $P$, i.e., $\Delta P=\frac{e}{2\pi} \int_0^\tau dt \int_\frac{-\pi}{a} ^\frac{\pi}{a} dk \sum_{n \in occu} f_n (k)$ being $f_n (k) = i \left[\frac{\partial}{\partial k} <\psi_nk (t)\vert \frac{\partial}{\partial t} \vert \psi_nk (t)> -\frac{\partial}{\partial t} <\psi_nk (t)\vert \frac{\partial}{\partial k} \vert \psi_nk (t)>\right]$ and under a gauge transformation of the electromagnetic potentials $A_\mu  {\rightarrow} A'_\mu= A_\mu + \partial_\mu \Lambda(x,t)$ the electron wave function transforms $\psi{\rightarrow} \psi=\psi e^{ie\frac{\Lambda}{\hbar}}$. The above exponential function needs to be single valued while the $\Lambda(x,t)$ doesn't. Thus we can write it as $\Lambda(x,t)=\frac{2\pi \mu \hbar \tau}{e\hbar \beta}=\mu\frac{2\pi\tau}{e\beta}$ where $\mu$ is a winding number quantizing the temperature. This allows to find the electric potential directly related with the temperature by 
\begin{equation}
V {\rightarrow} V'=V+\mu\frac{2\pi}{e}k_BT
\label{potential}
\end{equation}
which means that we have transformed the electric potential into another, plus a $\mu\frac{2\pi}{e\beta}$ thermal term. This turns out to be a fundamental result: the thermal energy appears quantized by the winding number being added to one electric potential under a gauge transformation. Notice that a Chern-Simons term was necessary since it is not gauge invariant and appears as surface actions. This allows to have, in non-Abelian Chern-Simons, the coupling constant $g$, directly associated to the temperature, related to the Chern-Simons action transforming $S_{CS} {\rightarrow} S'_{CS}=S_{CS}+\frac{2\pi \hbar^2\mu}{g^2} $ where $\mu$ is the topological mass.

Getting back to the topological electric potential equation (7), we can easy calculate the Seebeck coefficient $S=\mu \frac{2\pi}{e}k_B+\frac{2\pi}{e}\frac{\partial \mu}{\partial T} k_B T$ choosing $V'=0$. We can identify two contributions, the first, that comes purely from the temperature gradient in a topological branch, and the second, which takes into account the contribution due to a change in the topological index $\mu$, i.e, a jump between ramifications. As we will see, the first term is equivalent to that we find when we compute the surface Seebeck coefficient while the second, doesn't appear since there is only one ramification available on the surface. Once we have seen that the instanton solutions allow us to relate the electric potential $V$ with the temperature $T$ and hence the electric field $\textbf{E}$ with $\nabla T$ assuming no spatial homogeneity due to have topological bands, we can give a microscopic picture of how the thermal energy is invested in electricity. The main idea is that there is a creation (Fig. 3) of electron-hole Schwinger's pairs \cite{dumlu2011complex,dunne2005worldline,dunne2006worldline,D}, provided that the electric field is big enough. Calculating the critical electric field $E_c$, following Heisenberg-Euler \cite{heisenberg1936w}, we notice that the limit for obtaining a great number of real electron-hole pairs is 

\begin{equation}
E_c= \frac{m^ 2v_F^3}{e\hbar} \simeq 0.152 V/nm
\end{equation}
where we have considered that the energy gap is $0.21 eV$ and the Fermi velocity $v_F\simeq 6 \times 10^5 m/s$. This provides a critical electric field of almost ten orders of magnitude lower than the critical electric field in QED and with one equivalent temperature of $\zeta=1.74 \times 10^{-6} K nm^{-1}$, being these values accesible in these topological materials at so small scales as at hundredths of volt at distances of angstroms. In this way, we can rewrite the second term of the Seebeck coefficient considering that the dependence of $\mu$ with the temperature, as we show in equation (4), is represented by a Heaviside function, leading to the following expresion for the Seebeck

\begin{equation}
S= \frac{2\pi}{e}\mu k_B+\frac{2\pi}{e}\delta(\bar{p}) k_B T
\label{entropy}
\end{equation}
being $\bar{p}$ the different values where $\mu$ changes, that is, $0$, $\pm m/2$, $\pm m/4$ and where the local increase Fig. 4, represented by the second term, can be interpreted as the contribution originated in the creation of real electron-hole pairs. This equation 9 of the Seebeck coefficient has two terms, the first one quantize S in integers due to μ Chern-Simons topological mass, whereas the second corresponds to the variation of this mass respect the temperature T. The topological mass is quantized in steps of Heaviside respect to the temperature and therefore its derivative is equivalent to delta functions carrying the singularities. The relevant point is that the topological bands allow to find a new term for the Seebeck coefficient which can increase it depending of the value of temperature. But what is more important, these strings of singularities separate some regions from other thermally allowing a gradient of temperature ($\nabla T$) which produce an electric field $-\textbf{E}$ ($\nabla V$) in formula 7. This field $\vec{E}_2$ enables to create Schwinger's pairs that have different velocities depending of the level of the occupation states where they are situated respect to the Fermi level. Thus, at difference of what happens with metals, both effects of temperature on the electrons and holes can be quite different without destroying each other and the Seebeck coefficient would be also higher than a semimetal. It is fundamental to observe that without the topology ramifications of the topological insulators we would have an homogeneous crystal without thermoelectricity.

\begin{figure}[h]
\includegraphics[scale=0.85]{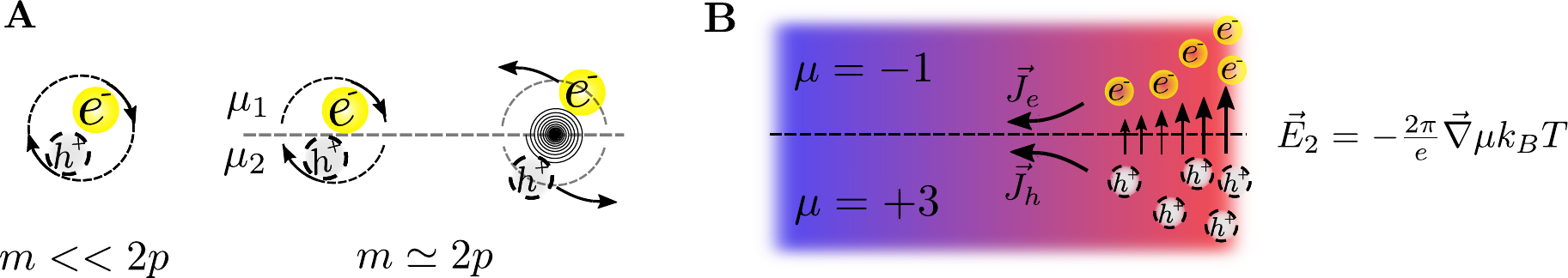}

\caption{\textbf{Particle hole creation close to the topological bands}. (\textbf{A}) Schematic illustration of a virtual electron-hole pair when the parameter $m$ is far away from the boundary between two topological branches, that is for example $2p$. When $m$ is close to these value the jump in the topological mass $\mu$ generates a potential difference that is able to break the vacuum and to generate a real electron-hole pair representing a new mechanism to transform thermal energy into electric in solid state physics. (\textbf{B}) Representation of electron and holes creation at the hot (red) side of the TI where $m\approx2p$. Electrons and holes take place on different branches, leadind to an electric field $\vec{E}_2$. Due to charge carriers are responsible to thermalize the material, this originates electron $J_e$ and holes $J_h$ currents where each type of carriers have different movilities since they below to diferent branches.}

\end{figure}

\begin{figure}[h]
\begin{center}
\includegraphics[scale=0.1]{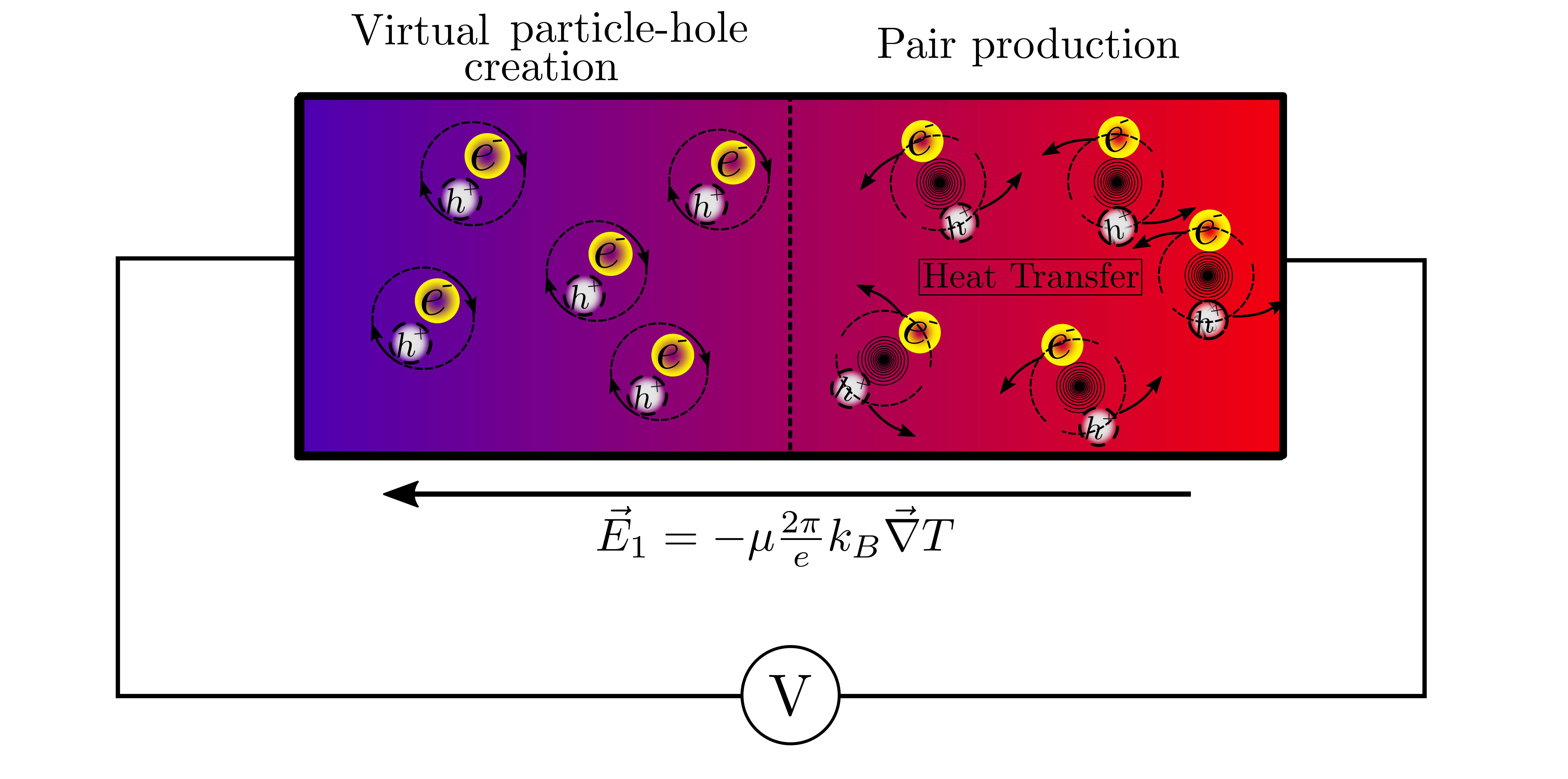}
\caption{\textbf{Non-Abelian topological contribution to the thermoelectric effect close to the U(1) electromagnetic surface}. Charge transport through the cold (blue) side originates a potential difference $V$ represented by the electric field $\vec{E}_1$. Pair production in the hot side (red) of the material leads to a local heat transfer, increasing the Seebeck coefficient.}
\end{center}
\end{figure}

Finally, let's calculate the topological contribution to the surface figure of merit in TI, limiting our calculations to (2+1) dimensions, where topology is determined through the first Chern number $\bar{n}$ and the winding number $n$. We start considering that, in a given direction $x$ or $y$, exists a temperature gradient in our TI in such a way that eigenstates evolve adiabatically. We can define then the Seebeck coefficient through the entropy  as $S=\frac{k_B}{e} \left(ln|\frac{\psi_b}{\psi_t}| + in\theta\right)$ where $\psi_b$  and $\psi_t$ are the final an initial states, $\theta$ is the angle between the states and $n$ is the winding number which can only take odd integers values, otherwise the topological Seebeck coefficient would give zero. In fact, we are representing the entropy of a Riemann surface. Notice that although the von Neumann entropy would give zero considering pure states, this is not true considering that these states are in entanglement due to their global topological properties. Actually there are many local Wannier bands, i.e., $\Psi(k,t)=\frac{1}{\sqrt{N}} \sum_i{\psi_i(k_i,t)}$, providing separate electrons which move coherently producing what is called Thouless charge pump\cite{Thouless} (see Fig. 5) and relates Berry phase with the electric polarization $P$ \cite{Resta},\cite{Switkes}. In fact, due to the previous definitions it is easy to see that $ln|\frac{\psi_b}{\psi_t}|=0$ and we can simplify the expression of the Seebeck coefficient for the TI as

\begin{equation}
       S=\frac{k_B}{e} in\theta=n\frac{k_B}{e} \int_C i<\psi(\textbf{r})|\nabla|\psi(\textbf{r})> d(\textbf{r})=\frac{\pi}{e} n \bar{n} k_B
\end{equation} 
being $\theta$ the angle directly associated to the Berry phase on the closed curve C. This leads to an expression very similar to that obtained in (4+1)D ,first term of equation (9), where the product $\mu w(U)$ has been transformed into $n\bar{n}$, where $\pi$ is the Berry phase of a non trivial material, $n$ counts the number of times we complete a cycle in a system (number of singularities) and $\bar{n}$ is the first Chern number which takes into account the whole topology on the Brillouin zone for our TI. For an adiabatical evolution of the eigenstates, we can go further in the physical interpretation of the Seebeck coefficient in TIs calculating its variation during one cycle ($n=1$) in the vector parameter
\begin{equation}
       \Delta S=\frac{k_B}{2\pi e} \int_0 ^\tau dt \int_\frac{-\pi}{a} ^\frac{\pi}{a} dk \sum_{n \in occu} \frac{\partial a_n (k)}{\partial t}
\end{equation} 
where a is the lattice constant considered in the Brillouin zone, $\tau$ is the period of time that takes a complete cycle in the parameter space, and $a_n (k)$ the Berry's connection in the k-space whose temporal derivative is taken on all the occuped states in the TI. Notice that $\tau$ will depend on the temperature gradient in the same way as it does on an external electric field which magnitude tell us  how fast the eigenstates evolution occur. Applying Stokes we can obtain $\Delta S$ as a function of the Berry's curvature $f_n(k)$ given an associated electric polarization $\Delta P$
\begin{equation}
       \Delta S= \frac{k_B}{ e^2} \Delta P
\end{equation} 
  where $\Delta P$ is \cite{Thou} 
	\begin{equation}
       \Delta P= \frac{e}{ 2 \pi} \int_0 ^\tau dt \int_\frac{-\pi}{a} ^\frac{\pi}{a} dk \sum_{n\in occu} f_n (k)
\end{equation}
and the curvature  $f_n (k) = i [(\frac{\partial}{\partial k} <\psi_nk (t) )|\frac{\partial}{\partial t} | \psi_nk (t)> - ((\frac{\partial}{\partial t} <\psi_nk (t) )|\frac{\partial}{\partial k} | \psi_nk (t)>]$. Finally, the change in the Seebeck coefficient turns out to be a very simple quantity
\begin{equation}
       \Delta S= \frac{k_B}{e^2} \bar{n} a e=  \bar{n} a \frac{k_B}{e}
\end{equation} 
being used the simplest expression of the polarization $\Delta P=\bar{n} ae$.

For the electronic thermal conductivity we consider the 2D density of states in the semimetal region as
\begin{equation}
 D(\xi)=2 \frac{1}{(2\pi)^2} \int \delta (\xi-\hbar v_F k) k dk = \frac{1}{\pi \hbar^ 2 v_F^2} \xi
\end{equation}
which allows us to obtain the electronic thermal conductivity $\kappa_e$ as
\begin{equation}
 \kappa_e = \frac{1}{2} \frac{\partial}{\partial T} \int d\xi D(\xi) \xi f(\xi) v_F l = \frac{3 \zeta (3)}{h} k_B^2 T
\end{equation}
Being $f(\xi)$ de Fermi-Dirac distribution function and where we have supposed a temperature dependent mean free path $l$. On the other hand, given that we have a ballistic regime for the electronic transport, its conductivity $\sigma_e$ appears given by the simple expression 
\begin{equation}
 \sigma_e = \bar{n} \frac{e^2}{h}
\end{equation}

\begin{figure}[h]
\begin{center}
\includegraphics[scale=0.1]{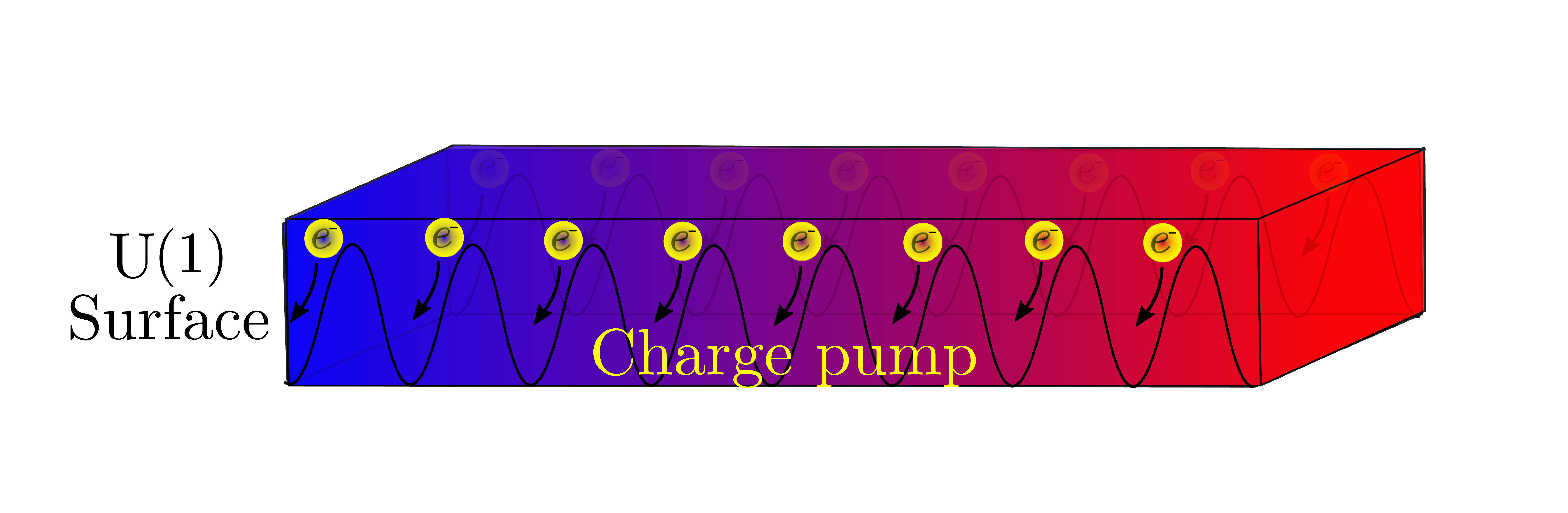}
\caption{\textbf{Electron charge pump in a quasi-one dimensional surface of a TI}. Electron transport takes place on the edge of the material with a quantized electrical conductivity $\sigma=\bar{n}e^2/h$}
\end{center}
\end{figure}

Therefore, in spite of using so different expressions than the ones of the metals, where a cuadratic dispersion equation is employed instead of the linear one of the semimetal, we obtain a good Wiedemann-Franz law yielded by 
\begin{equation}
 \frac{\kappa_e}{\sigma_e} = \frac{3 \zeta(3)}{\bar{n}} \left(\frac{k_B}{e}\right)^2 T=LT
\end{equation}
where the Lorenz L number is one constant, as it ought to be, but divided by a Chern number $\bar{n}$ which tell us that this expression is only valid within the context of the non-trivial topological materials that we have considered
\begin{equation}
 L = \frac{3 \zeta(3)}{\bar{n}} \left(\frac{k_B}{e}\right)^2 
\end{equation}
being $\zeta (3)$ the Riemann zeta function of dimension three (Apery's constant).
  
Finally we can calculate the figure of merit $Z$ for these topological insulators 

\begin{equation}
 Z=\frac{\sigma_e S^2}{\kappa_e} =\frac{S^2}{LT}
\end{equation}
where we are not considering the phononic part of the thermal conductivity \cite{holey,Peng}. In this way, the  dimensionless figure of merit turns out to be a simple expression
\begin{equation}
 ZT=n^2 \bar{n}^3\frac{\pi^2}{3\zeta(3)} 
\end{equation}
This is the extra pure topological figure of merit for the edge states, which is zero in the case of trivial topological materials. Although these conditions are quite ideal and transport constraints can diminish its efficiency under real physical features of each material, this result opens a great hope because it tell us how to improve highly the thermoelectricity associated to the topological materials. In the case of $Bi_2Te_3$ \cite{ven}, for the quantum numbers equal to one we obtain a value close to the one of its present maximum, i.e. $ZT=2.737$.

In summary, we have shown the relationship between topological insulators, as the family Bi$_2$Te$_3$ or topological related materials without time inversion protection as the Pb$_{1-x}$Sn$_x$Te, and their associated  thermoelectricity. We have also seen that the second Chern number obtained for the non-Abelian $SU(2)$ field leads to a thermal topological mass on a Chern-Simons action. This is equivalent to have a quantized  temperature working in a kind of topological bands that we define as the ramification branches using the Riemann-Hurwitz formula on an Euclidean spacetime where instanton solutions substitute Bloch oscillanting states. Physically what we have is a pumped charge between bands connected by the non-Abelian Berry phase within the insulator bulk at low temperature with an electromagnetic background field on the surface. Therefore, close to the surface we have only an Abelian $U(1)$ Chern-Simons term providing us with one transformation between electric and thermal energy because we have only one kind of states. Moreover we show that the Schwinger's electron-hole pairs, close to the topological bands, produce an increase of the Seebeck coefficient contributing to the transformation of thermal into electric energy  which is one of the key points of the model that we present in this paper. Finally we calculate a general expression to the dimensionless figure of merit in terms of the Chern number and winding number, getting a value that coincides quite well with the one experimentally measure for the Bi$_2$Te$_3$, doing zero its phononic thermal contribution. It is open for future a new class of topological materials using topological indices higher than one which can cross what is considered nowadays the efficient critical value of four for the ZT figure of merit changing the physical conditions suggesting in the presented model.

\providecommand{\noopsort}[1]{}\providecommand{\singleletter}[1]{#1}%

\section*{Acknowledgments}
 Thanks to the CESGA, to AEMAT ED431E 2018/08 and the MAT2016-80762-R project for financial support.

\clearpage

\end{document}